%% file: arXiv_ULTRA_Assignment.tex
\documentclass[sigconf,nonacm]{acmart}

\AtBeginDocument{%
  \providecommand\BibTeX{{%
    \normalfont B\kern-0.5em{\scshape i\kern-0.25em b}\kern-0.8em\TeX}}}

\setcopyright{acmcopyright}
\copyrightyear{2019}
\acmYear{2019}
\acmSubmissionID{186}
\acmDOI{10.1145/1122445.1122456}

\acmConference[Woodstock '19]{Woodstock '19: ACM Symposium on Neural
  Gaze Detection}{June 03--05, 2018}{Woodstock, NY}
\acmBooktitle{Woodstock '19: ACM Symposium on Neural Gaze Detection,
  June 03--05, 2019, Woodstock, NY}
\acmPrice{15.00}
\acmISBN{978-1-4503-9999-9/18/06}

\usepackage[linesnumbered,ruled,vlined,commentsnumbered]{algorithm2e}
\usepackage[percent]{overpic}
\usepackage{multirow}
\usepackage{pgfplots}

\pgfplotsset{compat=1.9}

\usetikzlibrary{shapes,arrows}

\input{kit_colors.tex}
\colorlet{refColor}{KITcyanblue}
\colorlet{ourColor}{KITgreen}
\colorlet{patColor}{KITred}
\colorlet{cycleColor}{KITorange}
\colorlet{assignmentColor}{KITseablue}
\colorlet{setupColor}{KITpalegreen}

\newcommand{\ultra}{ULTRA}
\newcommand{\thatIs}{i.e.\xspace}

\makeatletter
\renewcommand{\SetKwInOut}[2]{%
	\sbox\algocf@inoutbox{\KwSty{#2\algocf@typo:}}%
	\expandafter\ifx\csname InOutSizeDefined\endcsname\relax
	\newcommand\InOutSizeDefined{}\setlength{\inoutsize}{\wd\algocf@inoutbox}%
	\sbox\algocf@inoutbox{\parbox[t]{\inoutsize}{\KwSty{#2\algocf@typo\hfill:}}~}\setlength{\inoutindent}{\wd\algocf@inoutbox}%
	\else
	\ifdim\wd\algocf@inoutbox>\inoutsize%
	\setlength{\inoutsize}{\wd\algocf@inoutbox}%
	\sbox\algocf@inoutbox{\parbox[t]{\inoutsize}{\KwSty{#2\algocf@typo\hfill:}}~}\setlength{\inoutindent}{\wd\algocf@inoutbox}%
	\fi%
	\fi
	\algocf@newcommand{#1}[1]{%
		\ifthenelse{\boolean{algocf@hanginginout}}{\relax}{\algocf@seteveryparhanging{\relax}}%
		\ifthenelse{\boolean{algocf@inoutnumbered}}{\relax}{\algocf@seteveryparnl{\relax}}%
		{\let\\\algocf@newinout\hangindent=\inoutindent\hangafter=1\parbox[t]{\inoutsize}{\KwSty{#2\algocf@typo:}}~##1\par}%
		\algocf@linesnumbered
		\ifthenelse{\boolean{algocf@hanginginout}}{\relax}{\algocf@reseteveryparhanging}%
}}%
\renewcommand{\algocf@Vline}[1]{
	\strut\par\nointerlineskip
	\algocf@push{\skiprule}
	\hbox{\vrule\hspace{-0.4pt}%
		\vtop{\algocf@push{\skiptext}
			\vtop{\algocf@addskiptotal #1}\Hlne}}\vskip\skiphlne
	\algocf@pop{\skiprule}
	\nointerlineskip}
\makeatother

\SetKwInOut{Input}{Input}
\SetKwInOut{Output}{Output}
\DontPrintSemicolon
\newcommand{\blockEnd}{\vspace{-3.5pt}}

\SetKwFor{FOR}{for}{do}{}

\SetKwFor{FOREACH}{for each}{do}{}
\renewcommand{\ForEach}[2]{\FOREACH{#1}{#2\blockEnd}}

\SetKwFor{WHILE}{while}{do}{}

\SetKwIF{IF}{ELSEIF}{ELSE}{if}{then}{else if}{else}{}
\renewcommand{\If}[2]{\IF{#1}{#2\blockEnd}}

\newcommand{\positiverealnumbers}{\ensuremath{\mathbb R^+_0}\xspace}
\newcommand{\realnumbers}{\ensuremath{\mathbb R}\xspace}
\newcommand{\naturalnumbers}{\ensuremath{\mathbb N}\xspace}

\newcommand{\atime}{\ensuremath{\tau}\xspace}
\newcommand{\stops}{\ensuremath{\mathcal S}\xspace}
\newcommand{\astop}{\ensuremath{v}\xspace}

\newcommand{\connections}{\ensuremath{\mathcal C}\xspace}
\newcommand{\connection}{\ensuremath{c}\xspace}
\newcommand{\departurestop}{\ensuremath{\astop_\text{dep}}\xspace}
\newcommand{\arrivalstop}{\ensuremath{\astop_\text{arr}}\xspace}
\newcommand{\departuretime}{\ensuremath{\atime_\text{dep}}\xspace}

\newcommand{\arrivaltime}{\ensuremath{\atime_\text{arr}}\xspace}
\newcommand{\tripof}{\ensuremath{\text{trip}}\xspace}
\newcommand{\trips}{\ensuremath{\mathcal T}\xspace}
\newcommand{\atrip}{\ensuremath{\text{tr}}\xspace}
\newcommand{\buffertime}{\ensuremath{\atime_\text{buf}}\xspace}

\newcommand{\graph}{\ensuremath{G}\xspace}
\newcommand{\transferGraph}{\ensuremath{G}\xspace}
\newcommand{\footpathgraph}{\ensuremath{G}\xspace}
\newcommand{\vertices}{\ensuremath{\mathcal V}\xspace}
\newcommand{\vertex}{\ensuremath{v}\xspace}
\newcommand{\vertexa}{\ensuremath{v}\xspace}
\newcommand{\vertexb}{\ensuremath{w}\xspace}
\newcommand{\vertexc}{\ensuremath{x}\xspace}
\newcommand{\vertexd}{\ensuremath{y}\xspace}
\newcommand{\vertexe}{\ensuremath{z}\xspace}

\newcommand{\edges}{\ensuremath{\mathcal E}\xspace}

\newcommand{\walkingtime}{\ensuremath{\atime_\text{walk}}\xspace}
\newcommand{\apath}{\ensuremath{P}\xspace}

\newcommand{\demand}{\ensuremath{\mathcal{D}}\xspace}
\newcommand{\odpair}{\ensuremath{p}\xspace}
\newcommand{\origin}{\ensuremath{o}\xspace}
\newcommand{\destination}{\ensuremath{d}\xspace}

\newcommand{\distance}{\ensuremath{\operatorname{dist}}\xspace}

\newcommand{\journeys}{\ensuremath{\mathcal{J}}\xspace}
\newcommand{\journey}{\ensuremath{J}\xspace}
\newcommand{\profile}[2]{\ensuremath{f^{#1,#2}}\xspace}

\newcommand{\profilewait}[2]{\ensuremath{f_\text{wait}^{#1,#2}}\xspace}

\newcommand{\PAT}{\ensuremath{\arrivaltime^p}\xspace}
\newcommand{\pat}[1]{\ensuremath{\arrivaltime^p(#1)}\xspace}

\newcommand{\perceivedwalkingtime}{\ensuremath{\walkingtime^p}\xspace}

\newcommand{\passengermultiplier}{\ensuremath{\lambda_{\text{mul}}}\xspace}
\newcommand{\groupsize}{\ensuremath{\gamma}\xspace}
\newcommand{\transfercost}{\ensuremath{\lambda_{\text{trans}}}\xspace}
\newcommand{\waitingcost}{\ensuremath{\lambda_{\text{wait}}}\xspace}
\newcommand{\walkingcost}{\ensuremath{\lambda_{\text{walk}}}\xspace}
\newcommand{\buffercost}{\ensuremath{\lambda_{\text{buf}}}\xspace}

\newcommand{\delaytolerance}{\ensuremath{\lambda_{\Delta\text{max}}}\xspace}
\newcommand{\gain}{\ensuremath{g}\xspace}
\newcommand{\probability}{\ensuremath{P}\xspace}

\newcommand{\cs}{\extracolsep{\fill}}

\newcommand{\COMMENT}[2]{\ifnum0=#1\relax\else{}#2\fi}
\newcommand{\COMMENTSWITCH}[3]{\ifnum0=#1\relax{}#2\else{}#3\fi}

\newcommand{\extraSpaceBeforeTitle}{30pt}
\newcommand{\extraSpaceAfterTitle}{15pt}
\newcommand{\extraSpaceAfterAuthors}{35pt}

\begin{document}

\title[Efficient Computation of Multi-Modal Public Transit Traffic Assignments using ULTRA]{\vspace{\extraSpaceBeforeTitle}Efficient Computation of\\ Multi-Modal Public Transit Traffic Assignments\\ using ULTRA\\[\extraSpaceAfterTitle]}

\author{Jonas Sauer}
\email{jonas.sauer2@kit.edu}
\affiliation{%
  \institution{Karlsruhe Institute of Technology}
  \department{Department of Informatics}
  \city{Karlsruhe}
  \country{Germany}
}

\author{Dorothea Wagner}
\email{dorothea.wagner@kit.edu}
\affiliation{%
  \institution{Karlsruhe Institute of Technology}
  \department{Department of Informatics}
  \city{Karlsruhe}
  \country{Germany\\[\extraSpaceAfterAuthors]}
}

\author{Tobias Z\"undorf}
\email{zuendorf@kit.edu}
\affiliation{%
  \institution{Karlsruhe Institute of Technology}
  \department{Department of Informatics}
  \city{Karlsruhe}
  \country{Germany}
}

\begin{abstract}

We study the problem of computing public transit traffic assignments in a multi-modal setting:
Given a public transit timetable, an additional unrestricted transfer mode (in our case walking), and a set of origin-destination pairs, we aim to compute the utilization of every vehicle in the network.
While it has been shown that considering unrestricted transfers can significantly improve journeys, computing such journeys efficiently remains algorithmically challenging.
Since traffic assignments require the computation of millions of shortest paths, using a multi-modal network has previously not been feasible.
A recently proposed approach~(ULTRA) enables efficient algorithms with UnLimited TRAnsfers at the cost of a short preprocessing phase.
In this work we combine the ULTRA approach with a state-of-the-art assignment algorithm, making multi-modal assignments practical.
Careful algorithm engineering results in a new public transit traffic assignment algorithm that even outperforms the algorithm it builds upon, while enabling unlimited walking which has not been considered previously.
We conclude our work with an extensive evaluation of the algorithm, showing its versatility and efficiency.
On our real world instance, the algorithm computes over~15 million unique journeys in less than~17 seconds.
\vspace{5pt}
\end{abstract}



\maketitle

\vspace{5pt}
\section*{Funding}
This research was partially funded by the DFG under grant number: WA\,654123-2.

\section{Introduction}
\label{sec:introduction}

\begin{figure}[t]
  \centering
  \vspace{3pt}
  \begin{overpic}[width=\linewidth]{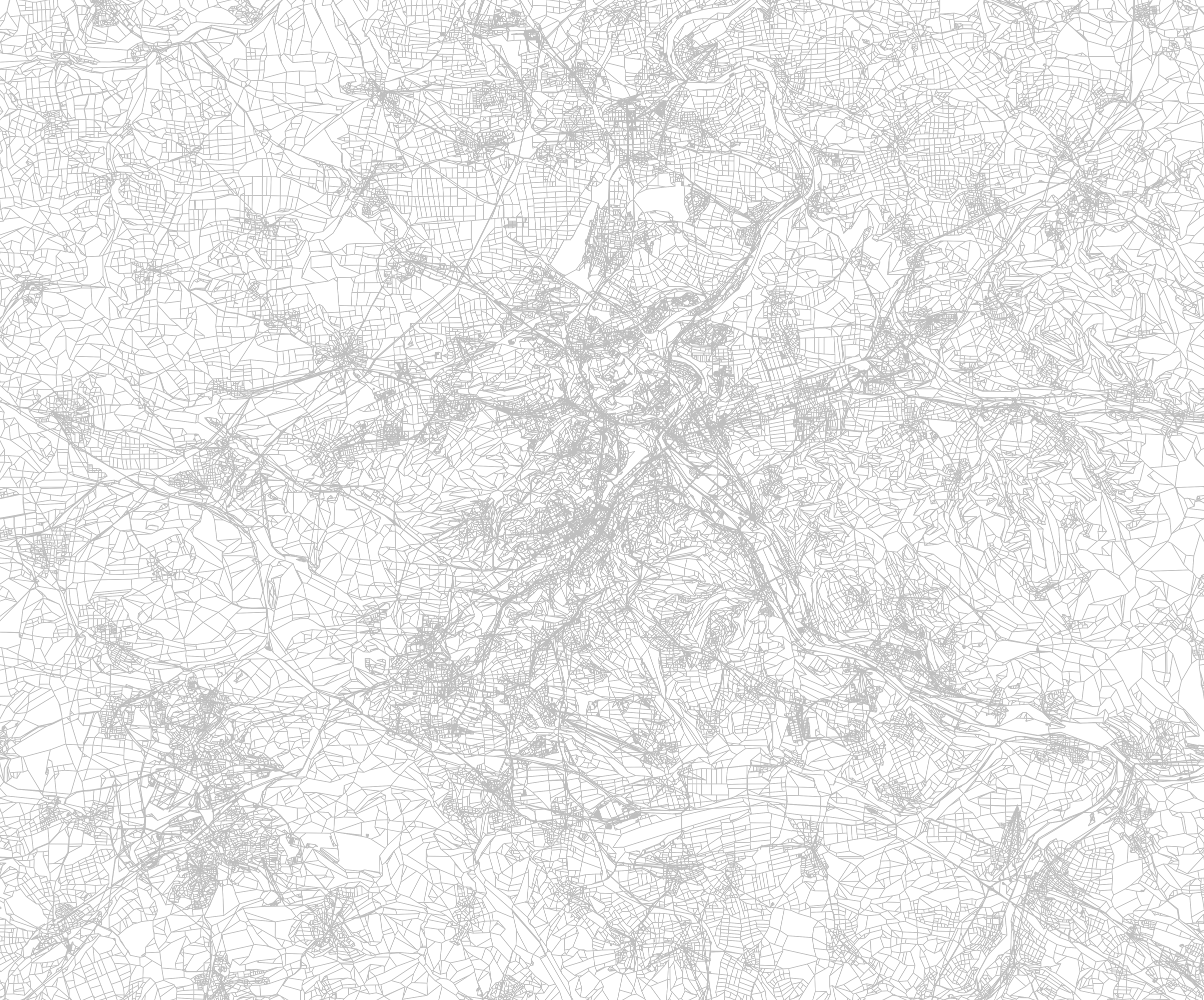}
     \put(0,0){\includegraphics[width=\linewidth,page=4]{fig/foreground2}}
  \end{overpic}
  \caption{
    Utilization of public transit vehicles between 8:30\,a.m.\ and 9:00\,a.m.\ in the Stuttgart area (Germany), as computed by our assignment algorithm.
    Vehicles used by only a few passengers are depicted using thin, green lines.
    As vehicles are used by more passengers, their line is gradually drawn thicker and their color changes to red.
    The transfer graph is depicted in gray.
  }
  \label{fig:visualization}
\end{figure}

Traffic assignments are an important tool for planning and analyzing transportation networks.
Efficient assignment algorithms allow to predict how new infrastructure could improve traffic flows, or to test the limits of existing networks, based on historic, empiric, or expected passenger demand data.
For this, the demand is given as a list of origin-destination pairs, where each pair is associated with a desired departure time.
A basic variant of the assignment problem then asks for the expected utilization of each vehicle~(\thatIs, the number of passengers using the vehicle) in the public transit network at each point in time.
A more intricate second variant additionally asks for a mapping from the origin-destination pairs onto actual journeys through the network that constitute the overall utilization of the vehicles.
Solving either of these problems efficiently requires both a fast route planning algorithm (in order to compute possible journeys for every origin-destination pair) and sophisticated decision models (in order to reflect adequately which journeys would be chosen by passengers in the real world).

Research on shortest path algorithms has made astounding progress throughout the past decade~\cite{bast2016}.
Highly efficient route planning algorithms in turn enable practicable assignment algorithms for decision models and settings of increasing complexity.
However, extending the underlying public transit network by other modes of transportation increases the difficulty of the problem significantly.
Nevertheless, considering walking without any limitation during journey planning can improve travel times substantially, as shown in~\cite{wagner2017,phan2019}.
The efficiency of public transit routing algorithms that can handle unrestricted transfers has recently been improved by a new \COMMENTSWITCH{0}{approach}{preprocessing technique} called ULTRA~\cite{ultra}, which promises to be easily adaptable for a wide range of route planning applications.
In this work we show how the public transit traffic assignment algorithm presented in~\cite{briem2017} can be adapted to work with ULTRA.
Moreover, we propose improvements to the algorithm that are independent from the used transfer graph, such that the resulting algorithm with unlimited transfers surpasses the performance of the~base~algorithm.

\subsection{Related Work}
\label{sec:relatedwork}

In general, traffic assignment problems can be subdivided into two mostly independent sub-problems.
First, computing high quality traffic assignments requires solving a classical route planning problem:
Given the origin, destination, and desired departure time of a passenger, find a set of all journeys that the passenger could reasonably use.
Second, a discrete choice model is required, which reflects the behavior of passengers in the real world and predicts the likelihood of each journey actually being used.
Both of these problems are widely studied on their own.
In the following we give a short overview of the most important results for each sub-problem.

\paragraph{Shortest Path Algorithms}
Regarding the route planning aspect of the traffic assignment problem, many efficient algorithms and speedup techniques have been developed in recent years.
A comprehensive overview of state-of-the-art route planning algorithms is given in~\cite{bast2016}.
Regarding the special case of route planning in public transit networks, many algorithms have been developed that exploit the special structure of timetables.
The RAPTOR algorithm~\cite{delling2014} is one of the first techniques based solely on an efficient timetable representation instead of a graph representation.
With Transfer Patterns~\cite{bast2010,bast2013}, a first approach that utilizes preprocessing in order to enable fast public transit queries was introduced.
A preprocessing technique called HypRAPTOR~\cite{delling2017} uses hypergraph partitioning to accelerate RAPTOR.
Using a data model that is focused on the trips made by the public transit vehicles, Trip-Based Routing~\cite{witt2015} and its accelerated version~\cite{witt2016} enable fast profile queries on public transit networks.
Yet another data model is used by the Connection Scan Algorithm~(CSA)~\cite{dibbelt2013}.
Here, the complete timetable is represented by a single sorted array of connections, which has to be scanned only once to solve earliest arrival or profile queries.
An accelerated version of this algorithm was proposed in~\cite{strasser2014}.
The approach of CSA was also used in the MEAT technique~\cite{dibbelt2014}, which takes possible delays into account during journey planning.
All algorithms mentioned so far were designed for pure public transit networks.
For multi-modal networks (\thatIs, networks \COMMENTSWITCH{0}{with}{that contain} at least one additional mode of transportation besides public transit) these algorithms are either not applicable or lose much of their efficiency.
However, a few algorithms have been proposed specifically for multi-modal routing:
MCR~\cite{delling2013} interleaves RAPTOR with Dijkstra's algorithm~\cite{dijkstra1959},
UCCH~\cite{dibbelt2015} adapts Contraction Hierarchies~\cite{geisberger2008} for multi-modal settings,
and HLRaptor and HLCSA~\cite{phan2019} combine Hub Labeling~\cite{abraham2011} with RAPTOR and CSA, respectively.
The most efficient approach to multi-modal route planning is currently ULTRA~\cite{ultra}.
This technique precomputes a small number of shortcuts that are sufficient to answer all queries correctly.
These shortcuts can then be easily incorporated into most one-to-one query algorithms.

\paragraph{Assignment Algorithms}
While route planning algorithms generally focus on finding journeys for a single user, traffic assignments aim at computing the overall traffic flow in a transportation network.
An overview of various traffic assignment models and algorithms can be found in~\cite{she85}.
In order to estimate the effects of congestion within parts of the network, equilibrium models are used, which are discussed in detail in~\cite{pat15}.
Similar to shortest path problems, the difficulty of traffic assignment problems increases when switching from road to public transit networks.
Thus, specialized traffic assignment techniques have been developed for public transit networks~\cite{pat15}.
An important part of most assignment algorithms are discrete choice models, which \COMMENT{0}{are used to }model the decision making process of the passengers mathematically.
Research on discrete choice models has a long history, as they are used in many areas, from operations research to economics.
An extensive overview of the most relevant choice models is given in~\cite{train2009}.
A \COMMENT{0}{very }natural approach to the integration of discrete choice models into public transit traffic assignment are sequential route choice models, where the computation of reasonable routes and the choice model are integrated.
Here, instead of computing complete routes from origin to destination and applying the choice model afterwards, routes are computed one leg at a time~\cite{gentile2006}.
Combining this approach with a suitable route planning algorithm (such as CSA) results in a very efficient traffic assignment algorithm~\cite{briem2017}.
The basic idea of the CSA-based assignment algorithm is to compute for each connection in the network a \COMMENT{0}{value called }\emph{perceived arrival time} (PAT) that reflects how useful the connection is in order to reach a certain destination.
A \COMMENTSWITCH{0}{decision}{discrete choice} model is then used to decide whether a passenger would actually use the connection based on the PAT.
Because of its efficiency, this approach has already been adapted for the travel demand simulation tool mobiTopp~\cite{briem2017b}.

\subsection{Our Contribution}
\label{sec:contribution}

In this work we present a novel public transit traffic assignment algorithm that considers not only the transit network but also unrestricted transfers.
For our new algorithm we combine the ULTRA preprocessing with the CSA-based assignment algorithm presented in~\cite{briem2017}.
Since ULTRA is designed for one-to-one queries while the assignment algorithm requires one-to-many queries, integration of the two techniques is not straightforward.
Therefore, we provide a detailed description of the steps necessary to adapt ULTRA for the assignment algorithm.
Furthermore, we present optimizations for the original assignment algorithm that increase the result accuracy and reduce computation time.
Finally, we demonstrate that the discrete choice model used by the algorithm can be exchanged with other choice models without affecting the efficiency.

\paragraph{Outline}
In Section~\ref{sec:preliminaries} we formalize the public transit traffic assignment problem as well as required network and demand models.
Furthermore, we provide a detailed introduction to the algorithms and techniques that our new approach builds upon in this section.
We proceed with the description of our new ULTRA-Assignment algorithm in Section~\ref{sec:assignment}.
Afterwards we present the results of an extensive experimental evaluation in Section~\ref{sec:experiments}, showing that our algorithm outperforms the state of the art.
We conclude our work with some final remarks in Section~\ref{sec:conclusion}.

\vspace{5.5pt}
\section{Preliminaries}
\label{sec:preliminaries}

In this section we introduce the notation used throughout the paper.
Furthermore, we present a short introduction of discrete choice models, the CSA-based assignment algorithm, and ULTRA, since our algorithm will build upon them.

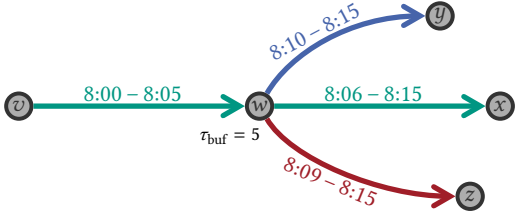
\begin{figure}[t]
    \centering
    \input{fig/transfer}
    \caption{Stop~\vertexb has a departure buffer time of $5$ minutes. It is possible to transfer from the green trip to the blue trip because the departure buffer time is observed, but not to the red trip. Staying in the green trip is also possible despite the departure buffer time not being observed.}
    \label{fig:buffertime}
\end{figure}

\subsection{Public Transit Routing}
We model a public transit network as a 4-tuple~$(\stops,\connections,\trips,\footpathgraph)$ consisting of a set of~\textit{stops}~$\stops$, a set of~\textit{connections}~$\connections$, a set of~\textit{trips}~$\trips$\!, and a directed, weighted~\textit{walking~graph}~$\footpathgraph=(\vertices,\edges)$.
Each \textit{connection}~$\connection\in\connections$ is a 5-tuple $(\departurestop(\connection), \arrivalstop(\connection), \departuretime(\connection), \arrivaltime(\connection), \tripof(\connection))$.
It represents a vehicle driving from a \textit{departure stop} $\departurestop(\connection) \in \stops$ to an arrival stop $\arrivalstop(\connection) \in \stops$ without halting in-between, departing at the \textit{departure time} $\departuretime(\connection)$ and arriving at the \textit{arrival time} $\arrivaltime(\connection)$.
A \textit{trip} $\atrip = (\connection_1, \dots, \connection_k) \in \trips$ is a sequence of connections served consecutively by a vehicle.
Every connection is part of exactly one trip, which is indicated by $\tripof(\connection)$.

The walking graph~$\footpathgraph=(\vertices,\edges)$ consists of a set of \textit{vertices}~$\vertices$ with~$\stops\subseteq\vertices$ and a set of \textit{edges}~$\edges\subseteq\vertices\times\vertices$.
Each edge~${(\vertexa,\vertexb)\in\edges}$ has a~\textit{walking time}~$\walkingtime(\vertexa,\vertexb)\in\naturalnumbers_0$ which defines the time needed to traverse the edge.
The definition of walking time can be extended naturally to paths in~$\footpathgraph$:
For a path~$\apath=(\vertex_1,\dots,\vertex_k)$, we define~$\walkingtime(\apath):=\sum_{i=2}^{k}\walkingtime(\vertex_{i-1},\vertex_i)$.
We denote the walking time of the shortest path in~$\footpathgraph$ from~$\vertexa$ to~$\vertexb$ by~$\distance(\vertexa,\vertexb)$.
Unlike in restricted walking scenarios, we require neither that $\footpathgraph$ is transitively closed, nor that if fulfills the triangle inequality.

Each stop~$\astop\in\stops$ is associated with a \emph{departure buffer time}~$\buffertime(\astop)$.
When arriving at $\astop$ with arrival time~$\atime$, passengers must observe the departure buffer time~$\buffertime(\astop)$ before they can enter a connection $\connection$ departing at~$\astop$, i.e., $\atime+\buffertime(\astop)\leq\departuretime(\connection)$ must hold.
The only exception to this is if the passengers arrived via the previous connection of $\tripof(\connection)$.
In this case, they can simply remain seated in the vehicle and do not have to enter it.
An example for the departure buffer time is shown in Figure \ref{fig:buffertime}.
Formally, transferring between two connections~$\connection_1$ and~$\connection_2$ is possible if~\mbox{$\tripof(\connection_1)=\tripof(\connection_2)$} and~$\connection_2$ immediately follows~$\connection_1$ in the trip, or if there is a path~${\apath=(\arrivalstop(\connection_1),\dots,\departurestop(\connection_2))}$ in the walking graph~$\footpathgraph$ connecting the stops~$\arrivalstop(\connection_1)$ and~$\departurestop(\connection_2)$ such that~ \mbox{$\arrivaltime(\connection_1)+\walkingtime(\apath)+\buffertime(\departurestop(\connection_2))\leq\departuretime(\connection_2)$} holds.
Note that in the special case of~$\arrivalstop(\connection_1)=\departurestop(\connection_2)$, $\apath$ may consist only of a single vertex.

\begin{figure}[t]
	\centering
    \input{fig/loop}
	\caption{An example of a footpath loop that can be used to bypass a minimum change time. Transferring from the green trip to the red trip at~\vertexb would normally not be possible due to the 5-minute change time, but taking the 3-minute footpath loop allows passengers to bypass it.}
    \label{fig:changetime}
\end{figure}
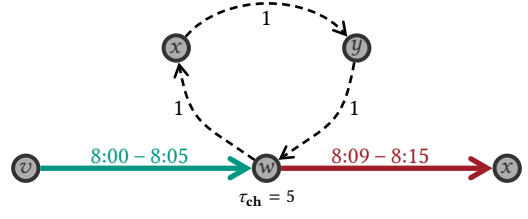

The departure buffer time models the time that is needed to walk to the platform where the vehicle departs and enter the vehicle, which may be considerable if the stop represents a large station.
Many other works, including the CSA-based assignment algorithm we want to build upon, instead consider \textit{minimum change times}~\mbox{\cite{delling2014,bast2010,dibbelt2013,briem2017b}}.
The minimum change time must only be observed when transferring between trips at the same stop, but not when entering a trip after arriving via a footpath.
In addition to arguably being unrealistic, this modeling choice also leads to problems when the walking graph~(including loop edges representing the minimum change time) is not required to be transitively closed.
Consider the situation depicted in Figure~\ref{fig:changetime}, where the minimum change time at a stop can be bypassed by taking a footpath that loops back to the stop.
In restricted walking scenarios, minimum change times are typically modeled as loop edges, and since the triangle inequality must be fulfilled, this situation is impossible.
To prevent this problem from occurring in the case of unrestricted walking, we use a departure time buffer time which must always be observed when entering a vehicle.
Note that loop edges in $\footpathgraph$ are allowed in our model, but they are superfluous since they can never improve the arrival time.
In contrast to the CSA-based assignment algorithm, ULTRA already considers departure buffer times.

We call the movement of a passenger from an origin vertex to a destination vertex through the public transit network a~\emph{journey}.
Formally a journey is an alternating sequence of the vehicles used by the passenger and paths in the transfer graph that allow for transferring between these vehicles.
By~$\arrivaltime(\journey)$ we denote the arrival time of the journey~$\journey$ at its destination~$\destination$.
If the journey uses at least one connection, then its arrival time is~$\arrivaltime(\connection)+\distance(\arrivalstop(\connection,\destination))$, where~$\connection$ is the last connection used within the journey.
If the journey uses no public transit vehicles, then its arrival time is simply the sum of the departure time at its origin and the distance from the origin to the destination.

The \emph{perceived arrival time}~(PAT)~$\pat{\journey}$ of a journey~$\journey$ is the sum of the actual arrival time~$\arrivaltime(\journey)$ and several penalties for inconveniences during the journey.
For each transfer between two vehicles (regardless of whether walking is required for the transfer or not), a transfer penalty~$\transfercost\in\positiverealnumbers$ is added to the perceived arrival time.
The time the passenger spends waiting for a vehicle to arrive is multiplied by a waiting penalty~$\waitingcost\in\positiverealnumbers$.
Similarly, the time the passenger spends walking is multiplied by a walking penalty~$\walkingcost\in\positiverealnumbers$.
Observe that the waiting and walking times contribute twice to the PAT: once in unweighted form as part of the actual arrival time, and then a second time as a weighted penalty.
Additionally, the perceived arrival time also takes into account the possibility of delays by incorporating alternative choices when a connection is delayed.
The precise formal definition of the perceived arrival time is given in~\cite{briem2017}.

A \textit{$\vertexa$-$\vertexb$-profile}~$\profile{\vertexa}{\vertexb}(\atime)$ between two vertices~$\vertexa,\vertexb\in\vertices$ is a function that maps each departure time~$\atime$ to the minimal perceived arrival time among all $\vertexa$-$\vertexb$-journeys that depart at~$\vertexa$ no earlier than~$\atime$.
If no feasible journey exists, we define~$\profile{\vertexa}{\vertexb}(\atime)$ as~$\infty$.
Note that every journey that involves taking at least one connection has a fixed departure time that depends solely on the first taken connection.
The profile can therefore be represented as a piecewise linear function where each segment has a fixed slope and ends with a breakpoint~$(\departuretime,\PAT)$, which represents a journey with the departure time $\departuretime$ and perceived arrival time $\PAT$.
Additionally, there may be at most one optimal journey which consists purely of walking from $\vertexa$ to $\vertexb$, which we represent with an additional value~$\walkingtime^p:=\walkingcost\cdot\distance(\vertexa,\vertexb)$.
Evaluating a profile at a time~$\atime$ is then done by returning the minimum of~$\walkingtime^p$ and the value of the function at $\atime$.

\subsection{Problem Statement}
A traffic assignment problem takes as input a public transit network and a \textit{demand}~\demand, which is a set of origin-destination pairs.
Each origin-destination pair $\odpair=(\origin,\destination)\in\demand$ is also associated with a desired departure time~$\departuretime(\odpair)$.
The pair~$\odpair=(\origin,\destination)$ represents a passenger who wants to travel from the \textit{origin vertex}~$\origin\in\vertices$ to the \textit{destination vertex}~$\destination\in\vertices$, starting at~$\departuretime(\odpair)$.
The objective of the traffic assignment problem is to assign each origin-destination pair~$\odpair=(\origin,\destination)\in\demand$ to a probability space consisting of journeys that satisfies the demand.
The demand is satisfied if each journey in the probability space departs at~$\origin$ no earlier than~$\departuretime(\odpair)$ and ends at~$\destination$.
The probabilities associated with each journey in the probability space should reflect the likelihood of a real passenger using the journey.
Summing the probabilities of all journeys containing a connection~\connection yields the \emph{utilization}~$u(\connection)$, which is the expected number of passengers using~\connection.

\subsection{Decision Models}
Most traffic assignment algorithms use discrete choice models in order to obtain the probability space for each origin-destination pair.
Given a \emph{choice set}, for example a set of possible journeys that end at the desired destination, a decision model can be used to determine the likelihood of each option in the choice set being used.
For this purpose every option is rated with a scalar value, called \emph{gain}, that reflects how useful the option is.
A decision model is then simply a function~$P\colon\realnumbers^n\to[0,1]^n$ that maps a vector containing the gains of all options onto a vector of probabilities.
Note that this requires~$\left\lVert\probability(x)\right\rVert_1=1$ for all~$x$.
We now briefly introduce some of the decision models that we want to use with our algorithm.

\paragraph{Logit}
The Logit model is a special case of a random utility model using a Generalized Extreme Value distribution Type-I.
This decision model is widely used since it has some useful properties and its corresponding probability function~$P_{\textit{logit}}(x)$ is comparatively simple.
Given~$n$ options with gains~$x_1,\dots,x_n$ and a tuning parameter~$\beta$, the Logit model defines the probability of choosing the~$i$-th option as
\[\probability_{\textit{logit}}(x)_i:=\frac{e^{\beta{}x_i}}{\sum_{j=1}^{n}e^{\beta{}x_j}}.\]

\paragraph{Kirchhoff}
In contrast to Logit, the Kirchhoff model is not rooted in discrete choice theory, but rather lends ideas from Kirchhoff's circuit laws.
The probability function~$\probability_{\textit{kirchhoff}}$ again depends on a tuning parameter~$\beta$ and is defined as
\[\probability_{\textit{kirchhoff}}(x)_i:=\frac{{x_i}^\beta}{\sum_{j=1}^{n}{x_j}^\beta}.\]

\paragraph{Linear}
The Linear decision model was proposed together with the CSA-based assignment algorithm in~\cite{briem2017}.
Given only two options, the probability of each option depends linearly on the difference between the options' gains, hence the name.
Let~$\bar{x}$ be the option with maximum gain and~$\delta$ the difference between the gains of the two best options.
Then the probability function~$\probability_{\textit{linear}}$ is defined as
\[\probability_{\textit{linear}}(x)_i:=\frac{\max(x_i, 2x_i-\bar{x}+\delta)}{\delta+\sum_{j=1}^{n}{x_j}^\beta}.\]

\subsection{CSA-Based Assignment}
The CSA-based assignment algorithm from~\cite{briem2017} computes a traffic assignment by simulating the movement of individual passengers through the public transit network.
The algorithm works by partitioning the origin-destination pairs based on their destination vertex and processing all pairs with the same destination vertex~$\destination$ simultaneously.
The assignment for the origin-destination pairs with destination~$\destination$ is done in three phases:~(1)~the PAT profile computation,~(2)~the actual assignment phase, and~(3)~an optional cycle removal phase.

The first phase computes a partial PAT profile~$\profilewait{\vertex}{\destination}$ for each vertex~$\vertex$ in the network, which is restricted to only consider journeys that start with a connection departing at~$\vertex$.
Thus,~$\profilewait{\vertex}{\destination}(\atime)$ is the minimum perceived arrival time for a passenger that is waiting at~$\vertex$ for the best vehicle departing after~$\atime$.
The CSA-based assignment algorithm computes these profiles with a single scan of the connection array in decreasing order of departure time.
During this scan the algorithm additionally computes three PATs for every connection that are later used to decide if passengers use the connection on their journey to the destination.

The second phase uses the PATs computed in the first phase to compute reasonable journeys with destination~$\destination$ and assign them to the origin-destination pairs.
This is done by simulating the movement of individual passengers through the network and recording the journey they take.
In order to obtain an actual probability space for each origin-destination pair~$\odpair=(\origin,\destination)$, and not just a single journey, the algorithm generates multiple passengers for~$\odpair$ and places them at~$\origin$.
The number of passengers per origin-destination pair is controlled by the \emph{passenger multiplier}~$\passengermultiplier\in\naturalnumbers$.
The passengers are then routed through the network using a variant of CSA, processing the connections in increasing order of departure time.
For each connection~$\connection$ the algorithm first decides whether passengers waiting at~$\departurestop(\connection)$ enter the connection in order to reach their destination.
Next the algorithm decides whether passengers that actually use the connection disembark at~$\arrivalstop(\connection)$.
Finally, the algorithm decides for passengers that disembark if they walk to another stop or keep waiting at~$\arrivalstop(\connection)$.
In all three cases the PATs of all available options for the passengers are already known, since they were computed in the first phase.
Given~$k$~options with~PATs~$\atime^p_1, \dots, \atime^p_k$, the gain~$\gain_i$ of the~$i$-th option is computed with the formula
\[\gain_i:=\max(0,\min_{j\neq{}i}(\atime^p_j)-\atime^p_i+\delaytolerance),\]
where~$\delaytolerance\in\positiverealnumbers$ is the \emph{delay tolerance}.
This causes options that differ from the best option by more than~\delaytolerance to have a gain of $0$, eliminating them from the choice set.
Using these gains, the probability for each option is computed with the Linear decision model.
A decision is then made for each passenger by choosing an option randomly according to the probabilities.
Doing this for each connection results in the passengers gradually moving towards the destination.\looseness=-1

As the passengers move randomly (but guided by the PATs) through the network, cycles can occur in their journeys.
If such cycles are undesired, they are removed in an optional third phase.
This is done by scanning all journeys recorded during the second phase for cycles and removing them.

The CSA-based algorithm obtains its efficiency from processing all passengers with the same destination at once.
Because of this, passengers coalesce as they get closer to their destination and the algorithm can evaluate the decision model for all of them at once.
Furthermore, in all three phases the computations for passengers with different destinations are completely independent of each other.
Thus, they can be performed in parallel.

\subsection{ULTRA}
The ULTRA approach~\cite{ultra} is based on the observation that vehicles in a public transit network are often coordinated.
Thus, long walking paths between vehicles are only rarely required, while long walking paths as the first or last leg of a journey occur quite frequently.
In order to exploit this, ULTRA uses a short preprocessing phase during which a shortcut graph~$\graph'=(\stops,\edges')$ containing all possible inter-vehicle transfers is computed.
Public transit algorithms can then find transfers between trips by scanning these shortcuts instead of the (much larger) original unrestricted transfer graph.
However, computing possible walking paths from the origin to the first stop and from the last stop to the destination still has to be done by the query algorithm.
The query algorithms proposed together with ULTRA do this efficiently by utilizing a specialized one-to-many shortest path algorithm called \mbox{Bucket-CH~\cite{knopp2007,geisberger2008}}.
This approach is only viable for solving one-to-one queries in the public transit network.
However, as the CSA-based assignment algorithm is based on an all-to-one CSA during the PAT computation and assignment phase, integrating ULTRA is not straightforward.

\section{ULTRA-Assignment Algorithm}
\label{sec:assignment}

In this section we give an overview of our ULTRA-Assignment algorithm, focusing on the differences to the original assignment algorithm introduced in~\cite{briem2017}.
In particular, we explain how departure buffer times and unrestricted walking can be integrated.
Furthermore, we outline how passengers representing the same origin-destination pair can be grouped to improve the running time and the accuracy of the results.
Pseudocode is given in Algorithm \ref{alg:assignment}.

\paragraph{Preprocessing for Zone Based Demand}
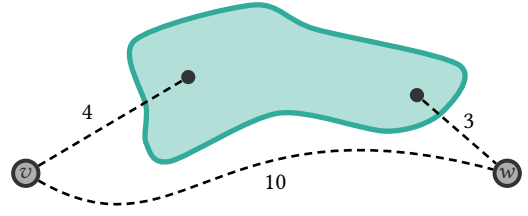
\begin{figure}[t]
    \centering
    \input{fig/zone}
    \caption{An example of a zone violating the triangle inequality. The vertices~\vertexa and~\vertexb are connected to the zone drawn in green, but the distances are calculated based on different endpoints within the zone. Adding the zone directly as a vertex would create an~\vertexa-\vertexb-path of length $7$, whereas the actual distance between~\vertexa and~\vertexb is $10$.}
    \label{fig:zone}
\end{figure}
The input to the assignment algorithm consists of a public transit network with an unrestricted walking graph and a demand.
Often, the demand data has a lower spatial resolution than the walking graph.
Origins and destinations are not supplied as precise locations, but rather as zones, which represent larger areas such as city districts.
In this case, distances between zones and nearby stops are supplied in the demand data.
Before the assignment algorithm can be executed, the zones have to be integrated into the public transit network.
However, simply adding vertices and edges for these zones to the walking graph may create new paths between stops that are too short and violate the triangle inequality.
This is because the zones represent regions with a non-zero expanse and the distances to nearby stops may be measured from any point within the region, not necessarily the center.
Two edges whose distances are measured from different endpoints within the region may form a path between stops that is too short, since it does not include the time needed to travel between the two endpoints.
An example of this is shown in Figure~\ref{fig:zone}.
To prevent this, we create two vertices for each zone: a source vertex that only has outgoing edges to stops and sink vertex that only has incoming edges from stops.
By not connecting the source and sink vertex, we prevent unwanted paths through zones from forming.

Once the zones have been integrated, we run the ULTRA preprocessing phase.
This involves computing a core graph for the shortcut computation, the shortcut computation itself, and computing the Bucket-CH data for the initial and final transfers.

\paragraph{Handling Departure Buffer Times}
A consequence of using ULTRA is that we must switch from the minimum transfer time model to the departure buffer time model.
Accordingly, passengers must observe the departure buffer time whenever a connection is entered, regardless of how it was reached.
In the original CSA-based assignment algorithm, the minimum transfer time was considered part of the waiting time and therefore the waiting penalty~\waitingcost was applied to it.
However, since the departure buffer time must always be observed, it may not be desirable to penalize it to the same degree as waiting, or at all.
Therefore we introduce a new buffer time penalty~$\buffercost\in\positiverealnumbers$ that may be different from the waiting penalty~\waitingcost.
Whenever a connection is entered, the departure buffer time multiplied by~\buffercost is added to the PAT.
Any time spent waiting before that, excluding the departure buffer time itself, is multiplied by~\waitingcost and added to the PAT.

\paragraph{Unrestricted Walking Using ULTRA}
\begin{algorithm}[t]
    \SetAlgoVlined
    \caption{\ultra-Assignment.}\label{alg:assignment}
	\Input{
		Public transit network~$(\stops,\connections,\trips,\transferGraph=(\vertices,\edges))$,\newline
		shortcut graph~$\transferGraph'=(\stops,\edges')$,\newline
		and demand~\demand
	}
	\Output{
		Utility~$u\colon\connections\to\positiverealnumbers$ for every connection,\newline
    	a set of journeys~$\journeys$ for each origin-destination pair
	}
    \BlankLine   
    \SetNlSty{bfseries}{\color{setupColor}}{}
    Let $O$ be the set of all origins with demand in~\demand\;
    Let $D$ be the set of all destinations with demand in~\demand\;
    \BlankLine
    \ForEach{$\origin\in{}O$\label{alg:buildDistanceLists}}{
        $N(\origin)\leftarrow\{(\astop,\distance(\origin,\astop))\mid\astop\in\stops\}$\label{alg:bucketCH}\tcp*[r]{Using Bucket-CH}
        Sort~$N(o)$ in ascending order of distance~$\distance(\origin,\cdot)$\label{alg:sortDistanceList}\;
    }
    \BlankLine
    Sort~\demand by destination\;
    Sort~\connections ascending by departure time\;    
    \SetNlSty{bfseries}{\color{black}}{}
    \ForEach{$\destination \in D$\label{alg:mainLoop}}{
        \SetNlSty{bfseries}{\color{setupColor}}{}
        Let~$\demand_\destination$ be the subset of~\demand with destination~\destination\;
        Sort $\demand_\destination$~by origin\;
        \SetNlSty{bfseries}{\color{patColor}}{}
        Compute PAT profiles from every stop to \destination\label{alg:patComputation}\;
        \SetNlSty{bfseries}{\color{assignmentColor}}{}
        \ForEach{$\odpair=(\origin,\destination)\in\demand_\destination$}{
            Generate passenger group~$g$ of size~\passengermultiplier for~\odpair\label{alg:generatePassengers}\;
            $\journeys\leftarrow\journeys\cup\{\journey_g=\{\}\}$\;
            Let $C$ be an empty choice set for~\odpair\;
            \ForEach{$(\astop,\distance(\origin,\astop))\in{}N(\origin)$}{\label{alg:evaluateDistances}
                $\departuretime\leftarrow\departuretime(\odpair)+\distance(\origin,\astop)$\label{alg:computeDepartureTime}\;
                $\perceivedwalkingtime\leftarrow\walkingcost\cdot\distance(\origin,\astop)$\label{alg:computePerceivedWalkingTime}\;
                $\bar\atime_{\text{arr}}^p\leftarrow\min\{\PAT\mid(\PAT,\cdot,\cdot)\in{}C\}+\delaytolerance$\;
                \If{$\departuretime+\perceivedwalkingtime>\bar\atime_{\text{arr}}^p$\label{alg:pruningCondition}}{\textbf{break}}
                $\PAT\leftarrow\profilewait{\astop}{\destination}(\departuretime+\buffertime(\astop)\hspace{-1pt})+\perceivedwalkingtime+\buffercost\cdot\buffertime(\astop)$\hspace{-5pt}\label{alg:evaluateProfile}\;
                $C\leftarrow{}C\cup\{(\PAT,\astop,\departuretime)\}$\;
            }
            Evaluate which choice from~$C$ the passengers use\label{alg:evaluateInitialWalking}\;
        }
        \ForEach{$\connection\in\connections$\label{alg:startAssignmentLoop}}{
            Evaluate if passengers waiting at~$\departurestop(\connection)$ enter~\connection\label{alg:evaluateEnter}\;
            $u(\connection)\leftarrow$ Number of passengers in~\connection\label{alg:computeUtilization}\;
            Add~\connection to journeys~$\journey_g$ of groups~$g$ in~\connection\label{alg:addToJourney}\;
            Evaluate if passengers using~\connection disembark at~$\arrivalstop(\connection)$\label{alg:evaluateDisembark}\;
            Evaluate if passengers at~$\arrivalstop(\connection)$ can transfer to~\destination\label{alg:evaluateDestination}\;
            Evaluate to which stop passengers at~$\arrivalstop(\connection)$ transfer\label{alg:evaluateTransfer}\;
        }
    }
    \SetNlSty{bfseries}{\color{cycleColor}}{}
    \ForEach{$\journey_g\in\journeys$\label{alg:startCycleCanceling}}{
        Remove cycles from~$\journey_g$\label{alg:endCycleCanceling}\;
    }
    \SetNlSty{bfseries}{\color{black}}{}
    \textbf{return} $u,\journeys$\;
\end{algorithm}

The original assignment algorithm used CSA variants with restricted walking in both the PAT computation phase and the assignment phase.
We extend the algorithm to unrestricted walking by replacing CSA with ULTRA-CSA in both phases.
This requires several changes.
The original PAT computation phase computed PAT profiles at every stop with a backward CSA search.
We replace this with a backward ULTRA-CSA search (line \ref{alg:patComputation}), using the shortcut graph for intermediate transfers and a backward Bucket-CH search from the destination~\destination for the final transfers.
These profiles exclude the initial transfers from the origin vertices, which are evaluated at the start of the assignment phase, when the passengers are generated and choose which stop they transfer to.

Evaluating the initial transfers constitutes the main algorithmic challenge.
The ULTRA technique was designed for one-to-one queries, where initial transfers can be computed with a single Bucket-CH query from the origin vertex.
In our case, there are typically multiple origin-destination pairs for the current destination vertex and therefore multiple origin vertices.
Thus we have to perform one Bucket-CH search for each origin vertex.
Once the initial transfers are computed, a further challenge is the efficient evaluation of the resulting transfer options.
In the restricted walking scenario, the choice set for initial and intermediate transfers was fairly small because passengers could only transfer to stops which were reachable via a direct edge.
This made it feasible to simply iterate over all outgoing edges, compute the PAT, gain, and probability for each reached stop, and then make a decision.
For the intermediate transfers, we can retain this approach in the unrestricted case by using the shortcut edges computed by ULTRA.
However, the shortcuts do not cover initial transfers.
When walking is unrestricted, almost all stops are reachable by initial walking from most origins.
Therefore it is no longer practical to explicitly collect all choices and compute their probabilities before making a decision.
In practice, however, the probability for the vast majority of options will be 0 because the required footpath is so long that the resulting PAT will exceed the delay tolerance~\delaytolerance.

To evaluate the initial transfers efficiently, we precompute the distances between the origin vertices and stops.
For each origin~\origin that occurs in the demand, we perform a Bucket-CH search from~\origin~to all stops (line \ref{alg:bucketCH}).
We store the distances from~\origin~to all reachable stops in a list of stop-distance tuples $(\astop,\distance(\origin,\astop))$, sorted in ascending order of distance (line~\ref{alg:sortDistanceList}).
After generating the passengers for an origin-destination pair $\odpair=(\origin,\destination)$ (line~\ref{alg:generatePassengers}), we iterate over the stop-distance tuples (line~\ref{alg:evaluateDistances}), compute the corresponding PATs and add them to the choice set~$C$.
For each tuple~$(\astop,\distance(\origin,\astop))$, we can compute the corresponding PAT by evaluating the profile~$\profilewait{\astop}{\destination}$ via binary search and adding the penalties for walking and the buffer time (line \ref{alg:evaluateProfile}).
Note that an option only has a non-zero gain and probability if its PAT does not exceed the PAT of the best option by more than~\delaytolerance.
To avoid iterating through the entire list of stop-distance tuples, we compute a lower bound for the PAT that increases monotonically with each tuple.
The lower bound consists of the departure time~\departuretime at $\astop$ (line \ref{alg:computeDepartureTime}) plus the perceived walking time~\perceivedwalkingtime (line \ref{alg:computePerceivedWalkingTime}).
Once this lower bound exceeds the best PAT found so far by more than~\delaytolerance, we can stop iterating through the list since all further options will have a probability of $0$ (line \ref{alg:pruningCondition}).

After we have collected all relevant options, we evaluate the choice set~$C$.
This involves computing the gain of each option, using a decision model to compute the probabilities and distributing the passengers to the stops according to the probabilities~(line~\ref{alg:evaluateInitialWalking}).
The rest of the assignment phase~(lines~\ref{alg:startAssignmentLoop}--\ref{alg:evaluateTransfer}) then continue~as usual, except that we use the shortcut graph for intermediate transfers and the final transfers computed by the Bucket-CH search from~\destination.
For each connection~\connection, four decisions are made:
First~it~is~decided which passengers waiting at~$\departurestop(\connection)$ enter~\connection~(line~\ref{alg:evaluateEnter}).
The utilization of~\connection is then calculated as the number of passengers using~\connection, including those that entered at~$\departurestop(\connection)$ and those that used a previous connection on the trip and remained seated~(line~\ref{alg:computeUtilization}).
Then, \connection is added to the journey of each passenger using it~(line~\ref{alg:addToJourney}).
Passengers in~\connection either decide to disembark at~$\arrivalstop(\connection)$ or remain in~$\tripof(\connection)$~(line~\ref{alg:evaluateDisembark}).
Those that disembark evaluate if they transfer directly to~\destination~(line~\ref{alg:evaluateDestination}).
If they do not, they choose a stop to transfer to next~(line~\ref{alg:evaluateTransfer}).
Each decision is made by using a decision model to compute the probabilities and distributing the passengers accordingly.
The cycle canceling phase~(lines~\ref{alg:startCycleCanceling} and~\ref{alg:endCycleCanceling}) remains~unchanged.

\paragraph{Grouping Multiplied Passengers}
The CSA-based assignment algorithm only approximates the unique solution defined by decision model, as the algorithm is based on the Monte Carlo method.
The accuracy of this approximation is primary influenced by the number of journeys that are sampled per origin-destination pair, which is controlled by the passenger multiplier~\passengermultiplier.
The original assignment algorithm generates~\passengermultiplier~copies of each passenger in the demand and then simulates the movement of all these copies independently.
While the gain computation and evaluation of the decision model can be shared among all passengers in the same location, each passenger still has to make an individual random decision and move accordingly.
This approach leads to redundant work because different copies of the same passenger will often make the same choices.
We solve this problem by grouping passengers that make the same choices together.

Instead of individual passengers, we now route \emph{passenger groups} through the public transit network.
The number of passengers in a group is indicated by the \emph{group size}~\groupsize.
At the start of the assignment phase, we generate one group of size~\passengermultiplier~for each origin-destination pair.
Previously, whenever we had to make a choice between options with probabilities $\probability_1, \dots, \probability_k$, we made an individual decision for each passenger by randomly picking an option according to the probabilities.
Now, when making a decision for a group of size~\groupsize, we split it into $k$ groups of sizes $\lfloor \groupsize \probability_1 \rfloor, \dots, \lfloor \groupsize \probability_k \rfloor$ and route each group according to the corresponding option.
Because the group sizes are rounded down, some of the original~\groupsize passengers may still be left over afterwards.
These passengers are still handled individually.
As before, we randomly choose an option for each passenger according to the probabilities and add the passenger to the corresponding group.
If the probability of an option is lower than $1 / \groupsize$ and none of the leftover passengers are assigned to it, the corresponding group has a size of $0$ and will be deleted.

When groups that represent different origin-destination pairs encounter each other at a stop, we do not merge them into a single large group.
While doing so would further reduce the computational effort, it would not allow us to reconstruct the journey that is assigned to each group in a straightforward manner.

The expected value for the share of passengers that are assigned to an option~$i$ is exactly~$\probability_i$, regardless of whether the passengers are grouped or not.
However, by grouping the passengers and splitting the groups according to the probabilities, a large portion of the assignment becomes deterministic.
Only the assignment of the leftover passengers created by rounding errors is still randomized.
Therefore, the computed utilization will not vary as strongly between different executions of the algorithm.

Instead of interpreting each unit in the simulation as a group of~\groupsize~passengers, we can also view it as a single passenger and interpret the group size as a fixed-point representation of the probability that the passenger will reach the current location, with~\passengermultiplier representing a probability of 1.
In this view, \passengermultiplier~is a parameter controlling the precision of the fixed-point representation and thereby the accuracy of the results.
If the precision was unlimited, rounding errors would no longer occur and the computed group sizes would conform exactly to the probabilities.
In this case, our algorithm would no longer be a Monte Carlo simulation but rather compute an exact solution of the assignment problem.

When decisions are made for each passenger individually, the computational effort of the algorithm is proportional to~\passengermultiplier.
With the grouped approach, the effort does not depend directly on~\passengermultiplier, but only on the number of group splits that are performed during the simulation.
This is limited by the number of feasible options.
Once the precision becomes high enough that each option with a non-zero probability is represented by a non-empty group, increasing the precision further may still improve the accuracy of the results, but it will not impact the running time.

\paragraph{Integrating Different Decision Models}
The original CSA-based assignment algorithm used the Linear decision model to compute probabilities out of the gains for each option.
However, any decision model that converts gains into probabilities without any additional information can be integrated directly into our assignment algorithm.
In this work, we implemented and evaluated the Logit and Kirchhoff model in addition to the Linear model.
Additionally, we implemented an \emph{optimal} decision model, which deterministically chooses the best option.
While this defies the purpose of a probabilistic assignment algorithm, it is nevertheless useful for comparing and analyzing the results.

\section{Experiments}
\label{sec:experiments}

\begin{table}[t]
    \center
    \caption{Sizes of the used public transit network, transfer graph, and demand.}
    \label{tbl:networks}
    \begin{tabular*}{\columnwidth}{@{\,}r@{\cs}r@{\cs}r@{\cs}r@{\cs}r@{\cs}r@{\,}}
        \toprule
        Vertices & Stops & Edges & Connections & Trips & Passengers \\
        \midrule
        1\,170\,198 & 13\,941 & 3\,710\,524 & 780\,042 & 47\,844 & 1\,249\,910 \\
        \bottomrule
    \end{tabular*}
\end{table}

All algorithms were implemented in C++17 compiled with GCC version 7.3.1 and optimization flag -O3.
All experiments were conducted on a machine with two 8-core Intel Xeon Skylake SP Gold 6144 CPUs clocked at~3.5\,GHz, 192\,GiB of DDR4-2666 RAM, and 24.75\,MiB of L3 cache.

\paragraph{Public Transit Network}
We conduct our experiments on a network representing the greater region of Stuttgart in Germany, as well as long distance services to nearby cities such as Frankfurt, Munich, and Basel.
The timetable data of this network and the accompanying demand data was also used in~\cite{briem2017} and was first introduced in~\cite{PTV11}.
The data covers the public transit service of one day, with the first connection departing at~0:39\,a.m.\ and the last connection arriving at~2:37\,a.m.\ of the following day.
We combine this timetable data with a transfer graph that was extracted from OpenStreetMap\footnote{\href{http://download.geofabrik.de/}{http://download.geofabrik.de/}}.
As we aimed for a transfer graph that represents walking between stops, we did not only extract streets, but also pedestrian zones and stairs from OpenStreetMap and computed walking times by assuming an average walking speed of~$4.0$\,km/h.
The transfer graph was combined with the timetable by identifying vertices of the graph with stops of the timetable if they were less than 5 meters apart from one another.
For stops that could not be identified with a vertex, we added transfer edges to their nearest vertex if the distance was less than 100 meters.
Nevertheless, after these steps~87 stops still remain without any connection to the transfer graph.
Finally we contracted vertices with degree one and two in the transfer graph, except for those vertices that were identified with a stop.
The size of the resulting combined network is shown in Table~\ref{tbl:networks}.

\subsection{Preprocessing}
Before the actual ULTRA-Assignment algorithm can be executed, we have to compute the required data structures in a preprocessing step.
As our query algorithm utilizes Bucket-CH queries, we have to compute a CH.
Additionally we need the ULTRA transfer shortcuts, which in turn require a core graph for their computation.
The CH was computed in~2:44\,min using a single thread and introduced~5\,469\,298~shortcuts.
The core graph was computed using CH with a limit of~16 for the average vertex degree within the core.
This resulted in a preprocessing time of~2:30\,min~(using one thread) and a core graph containing~25\,477 vertices and~407\,664 edges.
The core graph was then used for the ULTRA preprocessing, which took another~2:03\,min~(using~16 threads) and resulted in~74\,038 inter-vehicle shortcuts.

\subsection{ULTRA-Assignment}

The preprocessing step computing the ULTRA shortcuts and CH has to be repeated every time the used network changes.
However, the precomputed data structures can be reused for different demands.
In our evaluation of the assignment algorithm we used both the real demand data from~\cite{PTV11} and a demand of the same size but with origin and destination vertices picked uniformly at random.
For both demands we also evaluate how adding unrestricted walking transfers influences the resulting assignment.
An overview of the results is given in Table~\ref{tbl:results}.

In order to achieve reasonable precision in the result, we use a passenger multiplier of~100 for all experiments, unless stated otherwise.
Recall that this means we record~100 journeys for every origin-destination pair in the demand and thus compute the probability space of possible journeys for the demand pair accurately up to the second decimal place.
Furthermore, when comparing qualitative figures of the assignment (such as average travel time or number of used vehicles), we only consider origin-destination pairs that could be assigned to at least one journey in both networks, with and without unrestricted walking.
Using the original network, only~1\,209\,761 of the~1\,249\,910 origin-destination pairs could be assigned.
For all other pairs no feasible journey exists.
Using the unrestricted transfer graph, the number of assignable pairs increases to~1\,246\,337.
In both cases the reason for some pairs not being assignable is that our network represents only one day and the pairs have a departure time that is too late to reach their respective destination.

\begin{table}[t]
    \center
    \caption{
        Comparison of assignments with and without unrestricted transfers for real and random demand data.
        Results for networks with restricted transfers were obtained using the algorithm from~\cite{briem2017}.
        Results with unrestricted transfers use our new ULTRA-Assignment algorithm.
        Parallel execution times were measured using 16 cores.
        Figures concerning the quality of the computed assignment are averaged over all origin-destination pairs that were assigned to valid journeys both with and without unrestricted transfers.
    }
    \label{tbl:results}
    \begin{tabular*}{\columnwidth}{@{\,}l@{}r@{\cs}r@{\cs}r@{\cs}r@{\,}}
        \toprule
        & \multicolumn{2}{c}{Real demand}    & \multicolumn{2}{c}{\!\hspace{-5pt}Random demand\hspace{-5pt}\!}  \\
        \cmidrule(){2-3}                     \cmidrule(){4-5}
        Measured metric            & \hspace{5pt}Ref.\,\cite{briem2017} &         Our & \hspace{5pt}Ref.\,\cite{briem2017} &            Our\\
        \midrule
        Execution time (seq.) [s]  &                              299.6 &       181.9 &                              494.8 &          258.8\\
        Execution time (par.) [s]  &                               36.9 &        16.8 &                               52.9 &           19.7\vspace{3pt}\\
        Travel time [min]          &                               49.1 &        46.8 &                               91.2 &           82.2\\
        Walking time [min]         &                               22.2 &        22.2 &                               23.7 &           24.0\\
        Time in vehicle [min]      &                               21.8 &        20.7 &                               53.5 &           48.5\vspace{3pt}\\
        Connections per passenger  &                              10.76 &       10.19 &                              22.02 &          19.16\\
        Trips per passenger        &                               1.88 &        1.85 &                               3.06 &           2.88\\
        Journeys per passenger     &                               9.27 &       12.79 &                              13.43 &          17.85\\
        \bottomrule
    \end{tabular*}
\end{table}

Comparing the efficiency of our algorithm with the original algorithm from~\cite{briem2017}, we observe that despite solving a more difficult problem our algorithm outperforms the previous solution by a factor of about two regarding execution time (compare row 1 and 2 of Table~\ref{tbl:results}).
For either demand, the parallel version of the ULTRA-Assignment algorithm runs in below~20 seconds.
We also observe that assigning a random demand takes significantly longer than assigning the real demand.
A possible reason for this is that picking origin and destination vertices uniformly at random tends to produce long-distance demand pairs, while real demand contains more short-distance origin-destination pairs.
This assumption is backed by the different average travel times for both demands.
While a journey for a real origin-destination pair takes about~48 minutes, a journey for the random demand takes almost twice as long at about~80 to~90 minutes.
Additionally, a longer distance between origin and destination vertices tends to lead to a larger set of possible journeys.
Our measurements also confirm this correlation, with the number of assigned journeys increasing by about~50\% when switching from real demand to random demand.
A larger number of assigned journeys implies that more groups have to be split during the assignment phase, which directly affects the execution time of the algorithm and thus explains why the assignment for random demand takes longer.
For the real demand our algorithm assigns~12.79 journeys on average to each origin-destination pair.
Since the demand contains~1\,209\,761 pairs, this means that we can compute over~15 million individual journeys in less than~17 seconds.

\paragraph{Impact of Unrestricted Walking}
Compared to observations made in~\cite{phan2019} and~\cite{wagner2017}, adding unrestricted transfers to the network only has a small effect on the average travel times in our assignment.
For the real demand average travel times are only reduced by 4.6\% when switching from the original network to the unrestricted network.
However, results in~\cite{phan2019,wagner2017} were obtained on country-sized networks using random queries.
In contrast to country-sized networks, our network is much denser.
Furthermore, it can be assumed that the network was designed to match demand similar to our real demand data.
When considering random demand, the effect of unrestricted transfers already becomes more profound, as it reduces travel times by~9.8\% in this case.

\begin{figure}[t]
    \centering
    \input{fig/passengerMultiplier}
    \vspace{-10pt}
    \caption{
        Sequential execution time of assignment algorithms depending on the passenger multiplier.
        We compare ULTRA-Assignment to the algorithm presented in~\cite{briem2017}.
        For our algorithm we also report the execution time for several sub-phases of the algorithm.
        The measured running times are averaged over ten executions of the algorithm.
    }
    \label{fig:passengerMultiplier}
\end{figure}
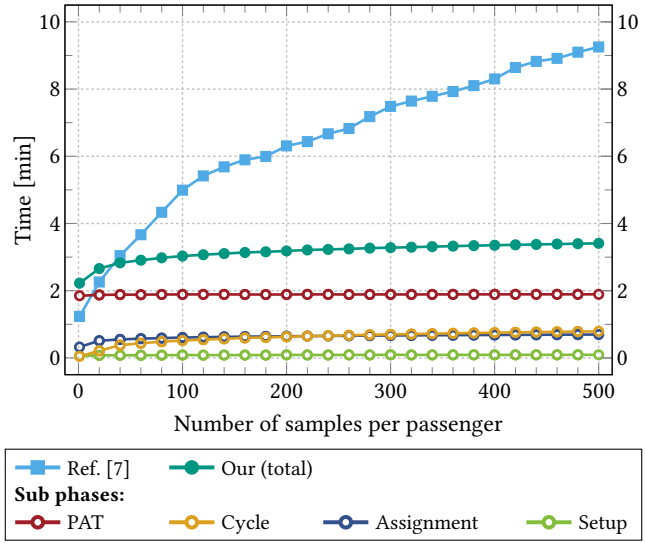

A visualization of the assignment computed by our algorithm for the real demand on the network with unrestricted transfers is shown in Figure~\ref{fig:visualization}.
The figure shows a small section from the center of the network, where every connection is colored according to the number of passengers assigned to it.
The visualization shows quite nicely the existence of a few main train lines (colored in red) that connect the outer parts of the network with the city center and are used by commuters in the morning.

\subsection{Passenger Multiplier}
The most important tuning parameter of our ULTRA-Assignment algorithm is the passenger multiplier~\passengermultiplier, which influences both the execution time of the algorithm and the accuracy of the computed assignment.
As stated before in Section~\ref{sec:assignment}, the effect of~\passengermultiplier on the result is quite direct, as the logarithm of~\passengermultiplier corresponds to the number of decimal places in the probability space that are computed exactly.
However, the effect of~\passengermultiplier on the execution time is not as clear.
We therefore evaluate the performance of our algorithm depending on the passenger multiplier in Figure~\ref{fig:passengerMultiplier}.
The plot shows that the total execution time increases with increasing~\passengermultiplier.
However, the impact of~\passengermultiplier on the total execution time decreases notably for high~\passengermultiplier.
The reason for this flattening of the curve is that the execution time of our algorithm does not depend directly on the passenger multiplier, but only on the number of group splits.
As more passengers are added, new groups are created less frequently since they represent options with increasingly small probabilities.
This result demonstrates the usefulness our new grouping approach during the assignment phase of the algorithm.

\begin{figure}[t]
    \centering
    \input{fig/decisionModels}
    \vspace{-10pt}
    \caption{
        Variation in the number of journeys per passenger and the execution time for different discrete choice models.
        We compare the Linear, Logit, and Kirchhoff models, each with different parameter settings.
        The used parameter values are annotated at the corresponding marker.
        For comparison we include the results of an \emph{optimal} decision model, where the journey with optimal gain is chosen deterministically.
        Running times are averaged over ten executions.
    }
    \label{fig:decisionModels}
\end{figure}
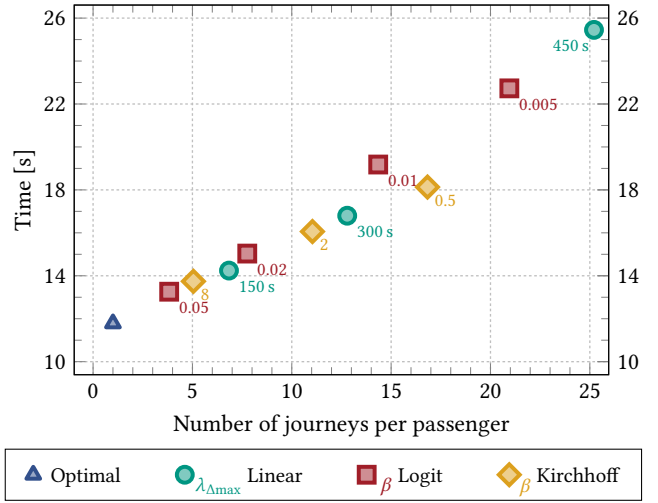

For comparison with the state of the art we also include the execution time from the algorithm from~\cite{briem2017}.
Direct comparison shows that the reference algorithm outperforms our algorithm for passenger multipliers below~50.
However, this is expected since our algorithm operates on a more complex network with unrestricted transfers.
On the other hand, for higher passenger multipliers our algorithm outperforms the previous approach.

Figure~\ref{fig:passengerMultiplier} additionally shows running times for the four sub-phases of our algorithm.
The colors of the four sub-phase curves correspond to the colors of the line numbers in Algorithm~\ref{alg:assignment}.
The plot shows that the most costly phase of our algorithm is the PAT computation phase.
This observation matches our expectations, as the PAT computation phase has to scan the complete multi-modal transportation network, including initial transfers.

\subsection{Decision Models}

With our last experiment we demonstrate the versatility of the ULTRA-Assignment algorithm by showing that our approach is compatible with a multitude of decision models.
We therefore evaluate the performance of our assignment algorithm combined with the Logit, Kirchhoff, and Linear decision model.
Additionally we test each decision model with different parameter settings.
The resulting execution times are shown in Figure~\ref{fig:decisionModels}.
Depending on the used decision model and its parameter settings, the execution time varies between~11 seconds and~26~seconds.
As before, the reason for this is primarily the number of different journeys that are assigned to each origin-destination pair, which correspond to the number of times that groups have to be split during the assignment phase.
To demonstrate this, we plotted the running time of the different models against the number of paths per passenger.
This plot clearly shows the linear correlation between the number of journeys per passenger and the execution time, confirming that our algorithm achieves the same efficiency for all tested decision models.

Besides execution time and number of assigned journeys, all decision models yield similar assignments.
The average travel time, for example, ranges from~46:28 minutes for the optimal assignment to~47:10 minutes for the Kirchhoff model with~$\beta=0.05$.
The reason for this is that additional suboptimal paths found by the algorithm are either only slight variations of the optimal path or have only a small probability and thus do not contribute much to the average.
Similar observations can be made for other quality metrics such as average number of vehicles used or average walking time.

\section{Conclusion}
\label{sec:conclusion}

We presented a new public transit traffic assignment algorithm that can handle unrestricted walking and is faster than previous approaches that were restricted to the pure public transit network.
We achieved this by integrating the novel ULTRA approach for handling unrestricted transfers into a state-of-the-art assignment algorithm.
By doing this, we developed the first one-to-many query algorithm that is able to use ULTRA shortcuts.
We proceeded with improving the overall performance of the assignment algorithm, so that we can compute an assignment for over~1.2 million origin-destination pairs in less than~17 seconds.
In a thorough experimental study we demonstrated the validity of our approach.
In particular we showed that our algorithm solves the assignment problem efficiently, regardless of the applied discrete choice model, the demand data, or the requested accuracy.

For future work, we would like to improve the overall quality of the computed assignments by integrating more complex journey choice models.
More sophisticated models could for example consider vehicle capacities and reduce the likelihood of assigning passengers to overcrowded vehicles.
Furthermore, it would be interesting to correlate the probabilities of journeys that overlap partially, for example if both use the same vehicle as a leg~of~the~journey.

\bibliographystyle{ACM-Reference-Format}
\bibliography{references}

\end{document}

%% file: kit_colors.tex
\usepackage{xcolor}
\definecolor{KITgreen}          {rgb}{0,    0.588,0.509}
\definecolor{KITgreen70}        {rgb}{0.3,  0.711,0.656}
\definecolor{KITgreen50}        {rgb}{0.5,  0.794,0.754}
\definecolor{KITgreen30}        {rgb}{0.7,  0.876,0.852}
\definecolor{KITgreen15}        {rgb}{0.85, 0.938,0.926}
\definecolor{KITblue}           {rgb}{0.274,0.392,0.666}
\definecolor{KITblue70}         {rgb}{0.492,0.574,0.766}
\definecolor{KITblue50}         {rgb}{0.637,0.696,0.833}
\definecolor{KITblue30}         {rgb}{0.782,0.817,0.9}
\definecolor{KITblue15}         {rgb}{0.891,0.908,0.95}
\definecolor{KITpalegreen}      {rgb}{0.509,0.745,0.235}
\definecolor{KITpalegreen70}    {rgb}{0.656,0.821,0.464}
\definecolor{KITpalegreen50}    {rgb}{0.754,0.872,0.617}
\definecolor{KITpalegreen30}    {rgb}{0.852,0.923,0.77}
\definecolor{KITpalegreen15}    {rgb}{0.926,0.961,0.885}
\definecolor{KITyellow}         {rgb}{0.98, 0.901,0.078}
\definecolor{KITyellow70}       {rgb}{0.986,0.931,0.354}
\definecolor{KITyellow50}       {rgb}{0.99, 0.95, 0.539}
\definecolor{KITyellow30}       {rgb}{0.994,0.97, 0.723}
\definecolor{KITyellow15}       {rgb}{0.997,0.985,0.861}
\definecolor{KITorange}         {rgb}{0.862,0.627,0.117}
\definecolor{KITorange70}       {rgb}{0.903,0.739,0.382}
\definecolor{KITorange50}       {rgb}{0.931,0.813,0.558}
\definecolor{KITorange30}       {rgb}{0.958,0.888,0.735}
\definecolor{KITorange15}       {rgb}{0.979,0.944,0.867}
\definecolor{KITbrown}          {rgb}{0.627,0.509,0.196}
\definecolor{KITbrown70}        {rgb}{0.739,0.656,0.437}
\definecolor{KITbrown50}        {rgb}{0.813,0.754,0.598}
\definecolor{KITbrown30}        {rgb}{0.888,0.852,0.758}
\definecolor{KITbrown15}        {rgb}{0.944,0.926,0.879}
\definecolor{KITred}            {rgb}{0.627,0.117,0.156}
\definecolor{KITred70}          {rgb}{0.739,0.382,0.409}
\definecolor{KITred50}          {rgb}{0.813,0.558,0.578}
\definecolor{KITred30}          {rgb}{0.888,0.735,0.747}
\definecolor{KITred15}          {rgb}{0.944,0.867,0.873}
\definecolor{KITlilac}          {rgb}{0.627,0,    0.47}
\definecolor{KITlilac70}        {rgb}{0.739,0.3,  0.629}
\definecolor{KITlilac50}        {rgb}{0.813,0.5,  0.735}
\definecolor{KITlilac30}        {rgb}{0.888,0.7,  0.841}
\definecolor{KITlilac15}        {rgb}{0.944,0.85, 0.92}
\definecolor{KITcyanblue}       {rgb}{0.313,0.666,0.901}
\definecolor{KITcyanblue70}     {rgb}{0.519,0.766,0.931}
\definecolor{KITcyanblue50}     {rgb}{0.656,0.833,0.95}
\definecolor{KITcyanblue30}     {rgb}{0.794,0.9,  0.97}
\definecolor{KITcyanblue15}     {rgb}{0.897,0.95, 0.985}
\definecolor{KITseablue}        {rgb}{0.196,0.313,0.549}
\definecolor{KITseablue70}      {rgb}{0.437,0.519,0.684}
\definecolor{KITseablue50}      {rgb}{0.598,0.656,0.774}
\definecolor{KITseablue30}      {rgb}{0.758,0.794,0.864}
\definecolor{KITseablue15}      {rgb}{0.879,0.897,0.932}
\definecolor{KITblack}          {rgb}{0,    0,    0}
\definecolor{KITblack90}        {rgb}{0.1,  0.1,  0.1}
\definecolor{KITblack80}        {rgb}{0.2,  0.2,  0.3}
\definecolor{KITblack75}        {rgb}{0.25, 0.25, 0.25}
\definecolor{KITblack70}        {rgb}{0.3,  0.3,  0.3}
\definecolor{KITblack60}        {rgb}{0.4,  0.4,  0.4}
\definecolor{KITblack50}        {rgb}{0.5,  0.5,  0.5}
\definecolor{KITblack40}        {rgb}{0.6,  0.6,  0.6}
\definecolor{KITblack30}        {rgb}{0.7,  0.7,  0.7}
\definecolor{KITblack25}        {rgb}{0.75, 0.75, 0.75}
\definecolor{KITblack20}        {rgb}{0.8,  0.8,  0.8}
\definecolor{KITblack10}        {rgb}{0.9,  0.9,  0.9}
\definecolor{KITwhite}          {rgb}{1,    1,    1}

%% file: fig/transfer.tex
\newcommand{\myScale}{0.8}
\newcommand{\edgeWidth}{2pt}
\newcommand{\ws}{\hphantom{--}}
\newcommand{\gs}{\hphantom{\tiny$\cdot$}}

\begin{tikzpicture}[line cap=butt,line join=round,>=angle 60,x=\myScale cm,y=\myScale cm]

\clip(-5,-1.8) rectangle (5,1.8);

\colorlet{nodeColor}{KITblack!80}

\tikzstyle{vertex}=[circle,line width=1.5pt,minimum size=0.1pt]
\node (u) at  (-4.0, 0.0) [vertex,draw=nodeColor!100,fill=nodeColor!40] {\gs};
\node (v) at  ( 0.0, 0.0) [vertex,draw=nodeColor!100,fill=nodeColor!40] {\gs};
\node (w) at  ( 4.0, 0.0) [vertex,draw=nodeColor!100,fill=nodeColor!40] {\gs};
\node (x) at  ( 3.0, 1.5) [vertex,draw=nodeColor!100,fill=nodeColor!40] {\gs};
\node (y) at  ( 3.5,-1.5) [vertex,draw=nodeColor!100,fill=nodeColor!40] {\gs};

\node (ut) at (u) [color=nodeColor] {\vertexa};
\node (vt) at (v) [color=nodeColor] {\vertexb};
\node (wt) at (w) [color=nodeColor] {\vertexc};
\node (xt) at (x) [color=nodeColor] {\vertexd};
\node (yt) at (y) [color=nodeColor] {\vertexe};

\node at (-0.5,-0.5) [] {\footnotesize$\buffertime=5$};

\draw [->,line width=\edgeWidth,color=KITgreen]  (u) -- (v) node [midway,above=-2pt] {8:00\,--\,8:05\ws};
\draw [->,line width=\edgeWidth,color=KITgreen]  (v) -- (w) node [midway,above=-2pt] {8:06\,--\,8:15\ws};
\draw [->,line width=\edgeWidth,color=KITblue]   (v) .. controls +(60:1) and +(180:1.5) .. (x) node [midway,above=-2pt,sloped] {8:10\,--\,8:15\ws};
\draw [->,line width=\edgeWidth,color=KITred]    (v) .. controls +(-60:1) and +(180:1.5) .. (y) node [midway,below=-2pt,sloped] {8:09\,--\,8:15\ws};

\end{tikzpicture}

%% file: fig/loop.tex
\newcommand{\myScale}{0.8}
\newcommand{\tripWidth}{2pt}
\newcommand{\edgeWidth}{1pt}
\newcommand{\ws}{\hphantom{--}}
\newcommand{\gs}{\hphantom{\tiny$\cdot$}}

\begin{tikzpicture}[line cap=butt,line join=round,>=angle 60,x=\myScale cm,y=\myScale cm]

\clip(-5,-1.8) rectangle (5,1.8);

\colorlet{nodeColor}{KITblack!80}
\colorlet{edgeColor}{KITblack}

\tikzstyle{vertex}=[circle,line width=1.5pt,minimum size=0.1pt]
\node (u) at  (-4.0, -1.0) [vertex,draw=nodeColor!100,fill=nodeColor!40] {\gs};
\node (v) at  ( 0.0, -1.0) [vertex,draw=nodeColor!100,fill=nodeColor!40] {\gs};
\node (w) at  ( 4.0, -1.0) [vertex,draw=nodeColor!100,fill=nodeColor!40] {\gs};
\node (x) at  ( -1.5, 1.0) [vertex,draw=nodeColor!100,fill=nodeColor!40] {\gs};
\node (y) at  ( 1.5,  1.0) [vertex,draw=nodeColor!100,fill=nodeColor!40] {\gs};

\node (ut) at (u) [color=nodeColor] {\vertexa};
\node (vt) at (v) [color=nodeColor] {\vertexb};
\node (wt) at (w) [color=nodeColor] {\vertexc};
\node (xt) at (x) [color=nodeColor] {$x$};
\node (yt) at (y) [color=nodeColor] {$y$};

\node at (0,-1.5) [] {\footnotesize$\atime_{\textbf{ch}}=5$};

\draw [->,line width=\tripWidth,color=KITgreen]  (u) -- (v) node [midway,above=-2pt] {8:00\,--\,8:05\ws};
\draw [->,line width=\tripWidth,color=KITred]    (v) -- (w) node [midway,above=-2pt] {8:09\,--\,8:15\ws};

\tikzstyle{edge}=[->,line width=\edgeWidth,color=edgeColor,densely dashed]
\draw [edge]   (v) .. controls  +(145:1.3) and  +(-80:1.3) .. (x) node [midway,left=3pt] {\raisebox{10pt}{1}};
\draw [edge]   (x) .. controls   +(50:1.2) and  +(130:1.2) .. (y) node [midway,below=-1pt] {1};
\draw [edge]   (y) .. controls +(-100:1.3) and   +(35:1.3) .. (v) node [midway,right=3pt] {\raisebox{10pt}{1}};

\end{tikzpicture}

%% file: fig/zone.tex
\newcommand{\myScale}{0.8}
\newcommand{\tripWidth}{2pt}
\newcommand{\edgeWidth}{1pt}
\newcommand{\ws}{\hphantom{--}}
\newcommand{\gs}{\hphantom{\tiny$\cdot$}}

\begin{tikzpicture}[line cap=butt,line join=round,>=angle 60,x=\myScale cm,y=\myScale cm]

\clip(-5,-1.6) rectangle (5,2);

\colorlet{nodeColor}{KITblack!80}
\colorlet{edgeColor}{KITblack}

\draw[line width=2pt,color=KITgreen!80,fill=KITgreen!30] plot [smooth cycle] coordinates {(0.8,1.4) (-0.2,1.8) (-2.2,1.2) (-2,0.1) (-2,-0.5) (-1.6,-0.8) (0.2,0) (2.0,-0.3) (2.9,0) (3.2,0.8) };

\tikzstyle{vertex}=[circle,line width=1.5pt,minimum size=0.1pt]
\node (u) at  (-4.0, -1.0) [vertex,draw=nodeColor!100,fill=nodeColor!40] {\gs};
\node (v) at  ( 4.0, -1.0) [vertex,draw=nodeColor!100,fill=nodeColor!40] {\gs};

\node (ut) at (u) [color=nodeColor] {\vertexa};
\node (vt) at (v) [color=nodeColor] {\vertexb};

\tikzstyle{dot}=[circle,fill,inner sep=1.5pt]
\node (zu) at  (-1.3, 0.6) [dot,fill=nodeColor!100] {\gs};
\node (zv) at  (2.5, 0.3)  [dot,fill=nodeColor!100] {\gs};

\draw [densely dashed,line width=\edgeWidth,color=edgeColor]  (u) .. controls +(-30:3) and +(165:5) .. (v) node [midway,below=-5pt] {\ws\hspace{40pt}10};
\draw [densely dashed,line width=\edgeWidth,color=edgeColor]  (u) -- (zu) node [midway,above left=-2pt] {4\ws};
\draw [densely dashed,line width=\edgeWidth,color=edgeColor]  (zv) -- (v) node [midway,above right=-2pt] {3\ws};

\end{tikzpicture}

%% file: fig/passengerMultiplier.tex
\tikzstyle{nicePlotA} = [line width=1pt,shorten <= 1.5pt,shorten >= 1.5pt,mark options={scale=0.8}]
\tikzstyle{nicePlotB} = [mark=o,line width=1pt,shorten <= 1.5pt,shorten >= 1.5pt,mark options={scale=0.8}]

\newcommand{\entry}[1]{\raisebox{-1.3pt}{\makebox[0.148\columnwidth][l]{#1}}}
\newcommand{\entryP}[2]{\raisebox{-1.3pt}{\makebox[#2\columnwidth][l]{#1}}}

\newcommand{\plotW}{1.045\columnwidth}
\newcommand{\plotH}{6.5cm}

\begin{tikzpicture}
\pgfplotsset{
   grid style = {dash pattern = on 1pt off 1pt, KITblack25,line width = 0.5pt}
}

\begin{axis}[
   height=\plotH,
   width=\plotW,
   xmin=-13,
   xmax=513,
   ymin=-30,
   ymax=630,
   xlabel={Number of samples per passenger},
   xtick={0, 100, 200, 300, 400, 500},
   minor x tick num={4},
   ylabel={Time [min]},
   ylabel style = {yshift=-5pt},
   ytick={0, 120, 240, 360, 480, 600},
   yticklabel=\pgfmathparse{\tick/60}${\pgfmathprintnumber{\pgfmathresult}}$\!,
   ytick pos=both,
   minor y tick num={1},
   grid=major,
   legend entries={\entry{Ref.~\cite{briem2017}}, \entry{Our (total)}, ~, ~,\hspace{-20pt}\textbf{Sub phases:}, ~, ~, ~, \entry{PAT}, \entry{Cycle}, \entryP{Assignment}{0.225}, \entryP{Setup}{0.078}},
   legend cell align=left,
   legend style={at={(-0.11,-0.2)},
   legend columns=4,
   anchor=north west,
   font=\small}
]

\addlegendimage{color=refColor,line width=1.5pt,mark=square*}
\addlegendimage{color=ourColor,line width=1.5pt,mark=*}
\addlegendimage{empty legend}
\addlegendimage{empty legend}
\addlegendimage{empty legend}
\addlegendimage{empty legend}
\addlegendimage{empty legend}
\addlegendimage{empty legend}
\addlegendimage{color=patColor,line width=1.5pt,mark=*,mark options={fill=white}} 
\addlegendimage{color=cycleColor,line width=1.5pt,mark=*,mark options={fill=white}} 
\addlegendimage{color=assignmentColor,line width=1.5pt,mark=*,mark options={fill=white}} 
\addlegendimage{color=setupColor,line width=1.5pt,mark=*,mark options={fill=white}} 

\addplot [color=refColor,nicePlotA,mark=square*] table {
   1 74.0657
   20 135.544
   40 182.633
   60 219.911
   80 260.056
   100 299.565
   120 324.965
   140 341.177
   160 353.766
   180 359.71
   200 378.584
   220 386.218
   240 400.168
   260 409.632
   280 430.853
   300 449.071
   320 458.554
   340 467.088
   360 475.637
   380 486.123
   400 498.191
   420 518.566
   440 529.42
   460 534.816
   480 545.889
   500 555.213
};

\addplot[color=ourColor,nicePlotA,mark=*] table[x=PassMult,y=Total,col sep=tab]{fig/passengerMultiplierResults.results};

\foreach \x in {1,...,25}{\addplot[color=setupColor,nicePlotB] table[x=PassMult,y=Setup,col sep=tab]{fig/passengerMultiplierResults\x.results};}
\foreach \x in {1,...,25}{\addplot[color=patColor,nicePlotB] table[x=PassMult,y=MEAT,col sep=tab]{fig/passengerMultiplierResults\x.results};}
\foreach \x in {1,...,25}{\addplot[color=assignmentColor,nicePlotB] table[x=PassMult,y=AssignmentInitWalking,col sep=tab]{fig/passengerMultiplierResults\x.results};}
\foreach \x in {1,...,25}{\addplot[color=cycleColor,nicePlotB] table[x=PassMult,y=CycleElim,col sep=tab]{fig/passengerMultiplierResults\x.results};}

\end{axis}

\begin{axis}[
   height=\plotH,
   width=\plotW,
   xmin=-13,
   xmax=513,
   ymin=-30,
   ymax=630,
   xmajorticks=false,
   ytick={0, 120, 240, 360, 480, 600},
   yticklabel=\pgfmathparse{\tick/60}\!${\pgfmathprintnumber{\pgfmathresult}}$,
   tick pos=right,
   ticklabel pos=right
]
\end{axis}

\clip (0,0) rectangle (1,5.16);

\end{tikzpicture}

%% file: fig/decisionModels.tex
\newcommand{\entryS}{\hspace{20.5pt}\!}
\newcommand{\entryA}[1]{\hspace{2pt}\raisebox{-1.3pt}{#1}\entryS}
\newcommand{\entryB}[3]{\raisebox{-5pt}{{\footnotesize\textcolor{#1}{#2}}}\hspace{2pt}\raisebox{-1.3pt}{#3}\entryS}
\newcommand{\entryC}[3]{\raisebox{-5pt}{{\footnotesize\textcolor{#1}{#2}}}\hspace{2pt}\raisebox{-1.3pt}{#3}\hspace{5pt}\!}

\newcommand{\plotLW}{1.5pt}
\newcommand{\plotS}{1.5}
\newcommand{\plotT}{1.4}

\newcommand{\plotW}{1.025\columnwidth}
\newcommand{\plotH}{6.5cm}

\begin{tikzpicture}
\pgfplotsset{
   grid style = {dash pattern = on 1pt off 1pt, KITblack25,line width = 0.5pt}
}

\begin{axis}[
   height=\plotH,
   width=\plotW,
   xmin=0,
   xmax=25,
   ymin=10,
   ymax=26,
   enlargelimits=0.038,
   xlabel={Number of journeys per passenger},
   xtick={0, 5, 10, 15, 20, 25},
   minor x tick num={4},
   ylabel={Time [s]},
   ylabel style={yshift=-3pt},
   ytick={10, 14, 18, 22, 26},
   minor y tick num={3},
   grid=major,
   legend entries={{\makebox(1,10){}}, \entryA{Optimal}, \entryB{KITgreen}{\!$\delaytolerance$}{Linear}, \entryB{KITred}{$\beta$}{Logit}, \entryC{KITorange}{\!\!$\beta$}{Kirchhoff}},
   legend cell align=left,
   legend style={at={(-0.13,-0.2),inner sep=100pt},
   legend columns=5,
   anchor=north west,
   font=\small}
]

\addlegendimage{empty legend}
\addlegendimage{color=KITseablue, line width=1.5pt, mark=triangle*, only marks, mark options={scale=\plotS, fill=KITseablue!50, line join=round}} 
\addlegendimage{color=KITgreen,   line width=1.5pt, mark=*,         only marks, mark options={scale=\plotS, fill=KITgreen!50}}                    
\addlegendimage{color=KITred,     line width=1.5pt, mark=square*,   only marks, mark options={scale=\plotS, fill=KITred!50}}                      
\addlegendimage{color=KITorange,  line width=1.5pt, mark=square*,   only marks, mark options={scale=\plotT, fill=KITorange!50, rotate=45}}        

\addplot [mark=triangle*,color=KITseablue,line width=\plotLW,mark options={scale=\plotS,fill=KITseablue!50,line join=round}] table { 
   x     y
   1     11.783
};

\addplot [
   mark=*,
   color=KITgreen,
   mark options={
   fill=KITgreen!50,
   scale=\plotS},
   line width=\plotLW,
   only marks,
   nodes near coords,
   nodes near coords align={anchor=north west},
   every node near coord/.append style={font=\footnotesize},
   point meta=explicit symbolic]
   table [meta=label] { 
   x      y       label
   6.84   14.242  150\,s
   12.79  16.797  300\,s
   25.21  25.451  \hspace{-20pt}450\,s
};

\addplot [
   mark=square*,
   color=KITred,
   mark options={
   fill=KITred!50,
   scale=\plotS},
   line width=\plotLW,
   only marks,
   nodes near coords,
   nodes near coords align={anchor=north west},
   every node near coord/.append style={font=\footnotesize},
   point meta=explicit symbolic]
   table [meta=label] { 
   x      y       label
   3.83   13.258  0.05
   7.76   15.034  0.02
   14.35  19.181  0.01
   20.95  22.728  0.005
};

\addplot [
   mark=square*,
   color=KITorange,
   mark options={
   fill=KITorange!50,
   scale=\plotT,
   rotate=45},
   line width=\plotLW,
   only marks,
   nodes near coords,
   nodes near coords align={anchor=north west,xshift=-1pt,yshift=1pt},
   every node near coord/.append style={font=\footnotesize},
   point meta=explicit symbolic]
   table [meta=label] { 
   x      y       label
   16.83  18.131  0.5
   11.04  16.06   2
   5.05   13.742  8
};

\end{axis}

\begin{axis}[
   height=\plotH,
   width=\plotW,
   xmin=0,
   xmax=25,
   ymin=10,
   ymax=26,
   enlargelimits=0.038,
   xmajorticks=false,
   ytick={10, 14, 18, 22, 26},
   tick pos=right,
   ticklabel pos=right
]
\end{axis}

\clip (0,0) rectangle (1,5.16);

\end{tikzpicture}